\documentclass[a4paper,11pt]{article}
\usepackage{pos}

\newcommand{\gsim}{\raisebox{-0.07cm   } {$\, \stackrel{>}{{\scriptstyle\sim}}\, $}}
\newcommand{\GeV}{\rm GeV}

\title{{\tiny \rm 
DESY-26-16, RISC Report Series 26--01, MPP--2026-17, PoS (RADCOR2025) 076
}\\
The complete three-loop unpolarized and polarized massive operator matrix elements and 
asymptotic Wilson coefficients}
\ShortTitle{The complete three-loop massive OMEs and Wilson coefficients}

\author[a]{J.~Ablinger}
\author[b]{A.~Behring}
\author*[c,d]{J.~Bl\"umlein}
\author[a]{A.~De Freitas}
\author[e]{A.~von Manteuffel}
\author[a]{C.~Schneider}
\author[f]{K.~Sch\"onwald}

\affiliation[a]{Johannes Kepler University Linz, Research Institute for Symbolic
Computation (RISC), Altenberger Stra\ss{}e 69, A-4040, Linz, Austria}

\affiliation[b]{Max-Planck-Institut f\"ur Physik
Boltzmannstra\ss{}e 8, 85748 Garching, Germany}

\affiliation[c]{Deutsches Elektronen-Synchrotron DESY, Platanenallee 6, 15738 Zeuthen, 
Germany}

\affiliation[d]{Institut f\"ur Theoretische Physik III, IV, TU Dortmund, Otto-Hahn
Stra\ss{}e 4, 44227 Dortmund, Germany}

\affiliation[e]{Institut f\"ur Theoretische Physik, Universit\"at Regensburg, 93040
Regensburg, Germany}

\affiliation[f]{CERN, Theoretical Physics Department,
CH-1211 Geneva 23, Switzerland}

\emailAdd{Jakab@gmx.at}
\emailAdd{arnd.behring@desy.de}
\emailAdd{Johannes.Bluemlein@desy.de}
\emailAdd{abilio.de.freitas@desy.de}
\emailAdd{manteuffel@ur.de}
\emailAdd{Carsten.Schneider@risc.jku.at}
\emailAdd{kay.schonwald@cern.ch}

\abstract{We report on the three-loop unpolarized and polarized massive operator matrix 
elements, with single- and two-mass corrections, and the associated deep-inelastic 
massive Wilson coefficients in the region $Q^2 \gg m_Q^2$, the calculation of which has 
been completed recently. We also provide fast and precise numerical representations of
the massless Wilson coefficients, splitting functions to tree-loop order, and target-mass 
corrections in $x$-space well suited for QCD-fitting codes.}

\FullConference{17th International Symposium on Radiative Corrections: 
Applications of Quantum Field Theory to Phenomenology (RADCOR2025)\\
5-10 October 2025\\
Puri, India\\}

\begin{document}
\maketitle

\section{Introduction}
\label{sec:1}

\vspace*{1mm}
\noindent
The precision measurement of the strong coupling constant $a_s(M_Z^2) = \alpha_s(M_Z^2)/(4 \pi)$
\cite{Bethke:2011tr,Moch:2014tta,Alekhin:2016evh,dEnterria:2022hzv}
and the mass of the charm quark $m_c$ \cite{Alekhin:2012vu} using deep-inelastic scattering data 
needs to  be performed  under well-defined experimental conditions.
Only for the twist--2 contributions, sufficiently precise calculations were performed. 
The higher
twist corrections have therefore to be removed by kinematic cuts
\cite{Blumlein:2006be,Blumlein:2008kz,Blumlein:2012se,Alekhin:2012ig}. At an intended precision of 
$O(1\%)$ and 
better, this requires to refer to data $Q^2 \gsim 20~\GeV^2$ and $W^2 = Q^2 (1-x)/x \gsim 
15~\GeV^2$
for the virtuality and the hadronic mass squared in the deep-inelastic scattering process.
This is also the region for which it has been shown in Ref.~\cite{Buza:1995ie} 
that the asymptotic heavy-flavor corrections
for the structure function $F_2(x,Q^2)$ at $O(a_s^2)$ agree at the level of $O(1\%)$ with the
full phase-space calculation in Refs.~\cite{Laenen:1992zk,Laenen:1992xs,Riemersma:1994hv}.

In the asymptotic region, analytic results can be calculated at $O(a_s^3)$ in the single- and 
two-mass cases. Here, simple alphabet structures ruling the massless case, such as 
harmonic sums 
\cite{Vermaseren:1998uu,Blumlein:1998if}, are not sufficient anymore. The solution spaces are spanned 
by generalized harmonic sums  \cite{Moch:2001zr,Ablinger:2013cf},
cyclotomic harmonic sums \cite{Ablinger:2011te},
nested binomial harmonic sums \cite{Ablinger:2014bra}, iterated integrals over letters implied by 
quadratic forms \cite{Ablinger:2021fnc},
and $_2F_1$- or complete elliptic letters (which are modular forms) \cite{Ablinger:2017bjx}.

Even more involved structures appear in the two-mass case. Of course, one can derive
the corresponding iterated-noniterative integrals \cite{JB2016}.
In the end one needs, however, fast and precise numerical implementations.\footnote{
Even for the numerical quantification of harmonic polylogarithms \cite{Remiddi:1999ew}
over a simple alphabet \cite{Blumlein:2003gb} of only three (non-singular)
letters, efficient numerical implementations have to be found 
\cite{Gehrmann:2001pz,Ablinger:2018sat,Vollinga:2004sn},
consisting of the  Bernoulli-method or H\"older convolution \cite{Borwein:1999js}, despite having a 
formal representation as iterative integrals.} 
For this purpose, the analytic method described in Ref.~\cite{Behring:2023rlq}
is very well suited.
It is based on overlapping deep local analytic expansions of logarithmically modulated 
Taylor expansions, allowing to reach any precision.

Analytic results in the full phase space were computed to $O(a_s^2)$ for the non-singlet and 
pure-singlet contributions both in the unpolarized and polarized cases in 
Refs.~\cite{Buza:1995ie,Buza:1996xr,Blumlein:2016xcy,Blumlein:2019qze,Blumlein:2019zux}.
This is also possible for the gluonic cases, at the expense of alphabets containing 
higher
transcendental letters. The non-singlet contributions at $O(a_s^2)$ already contain 
root-valued 
letters \cite{Ablinger:2014bra} and the iterated integrals in the pure-singlet case 
contain 
incomplete elliptic 
integrals \cite{Blumlein:2019qze,Blumlein:2019zux}. In Ref.~\cite{Alekhin:2003ev} a 
complete
Mellin-space implementation of the $O(a_s^2)$ results 
\cite{Laenen:1992zk,Laenen:1992xs,Riemersma:1994hv} has been derived for $N \in \mathbb{C}$ at high 
precision for the use in Mellin-space QCD analysis programs. 
The corresponding phase-space integrals at 
$O(a_s^3)$ will contain 
even more of these structures, and imply various thresholds and pseudo-thresholds to be dealt with 
in their numerical quantification. Furthermore, in part of the contributions one faces a three-scale 
problem due to the presence of both the charm and bottom quark masses.

In the following we will give a survey on the quantitative three-loop results of the 
heavy-flavor corrections to deep-inelastic scattering for larger virtualities both in the 
unpolarized and 
polarized case to three-loop order. We describe the computation methods used in 
Section~\ref{sec:2}. 
A survey on the single-mass corrections is given in Section~\ref{sec:3}, and the two-mass 
corrections 
are described in Section~\ref{sec:4}. In Section~\ref{sec:5}, we discuss
the implications for the structure function $F_2(x,Q^2)$. In 
Refs.~\cite{Ablinger:2025joi,Ablinger:2025awb} we also presented fast and precise numerical 
implementations for all unpolarized and polarized single-mass operator matrix elements (OMEs) and
asymptotic massive Wilson coefficients as public codes. In Section~\ref{sec:6}, 
we supplement 
these 
for the massless three-loop Wilson coefficients and splitting functions in the same form, for 
convenient use for $x$-space QCD-analysis codes. Moreover, we add a numerical 
implementation for the 
target-mass corrections. Section~\ref{sec:7} contains the conclusions.
\section{The Computation Methods}
\label{sec:2}

\vspace*{1mm}
\noindent
The massive three-loop operator matrix elements $A_{ij}$ are propagator integrals containing 
local operator insertions \cite{YND,Bierenbaum:2009mv}.  Their renormalization includes mass-,
coupling constant-, and
operator-renormalization, as well as the subtraction of the collinear singularities of massless 
sub-graphs, see Ref.~\cite{Bierenbaum:2009mv} in the single-mass case and 
Refs.~\cite{Ablinger:2017err,Ablinger:2025nnq} in the 
two-mass 
case. The factorization relations valid for virtualities $Q^2 \gg m_Q^2$, with $m_Q$ the 
heavy-quark 
mass, cf.~Refs.~\cite{Buza:1996wv,Ablinger:2025joi}, imply the representation of the 
heavy-flavor 
Wilson coefficients in terms of polynomials of respective pieces of the massive OMEs and the massless 
Wilson coefficients \cite{Vermaseren:2005qc,Moch:2008fj,Blumlein:2022gpp} in Mellin space in this 
kinematic region.
 
The calculation of the contributing Feynman diagrams from their generation to the reduction to the 
master integrals has been described in Ref.~\cite{Ablinger:2023ahe}. We use the packages {\tt 
QGRAF, Form, Color} and {\tt Reduze~2} 
\cite{Nogueira:1991ex,Vermaseren:2000nd,Tentyukov:2007mu,vanRitbergen:1998pn,Studerus:2009ye,
vonManteuffel:2012np}. In the polarized
case we compute the OME in the Larin scheme \cite{Larin:1993tq}. 

For the calculation of the different diagram classes, different calculation methods were 
used.
The simpler topologies can be computed directly using the method of (generalized) hypergeometric
functions \cite{HAMBERG,Bierenbaum:2007qe,Ablinger:2012qm,Ablinger:2015tua}, paired with 
direct summation methods \cite{Bierenbaum:2008yu,Ablinger:2010ty}. Here algorithms of difference ring 
theory were used \cite{Karr:1981,Bron:00,Schneider:01,Schneider:04a,Schneider:05a,Schneider:05b,
Schneider:07d,Schneider:2009rcr,Schneider:10c,Schneider:15a,Schneider:08d,Schneider:2017}.

In the more involved cases, the spanning functions both in Mellin- and $x$-space 
were not known a priori, but had to be derived algorithmically. Here the method of 
arbitrary high 
Mellin moments \cite{Blumlein:2017dxp} is essential, since it allows to find the input required to 
obtain the corresponding recursions from a 
finite number of moments. These are determined using guessing methods
\cite{GUESS,Blumlein:2009tj,SAGE,GSAGE}.
Next, it has to be analyzed by using the
package {\tt Sigma} \cite{SIG1,SIG2} whether the recursion is first-order-factorizing or 
not.
In the former case, one finds sum-product representations of the solution. 
This has to be the case for the {\it whole} physical set of Mellin-moments, 
cf.~Ref.~\cite{Blumlein:1996vs}.
In the case of the OMEs $A_{gg,Q}^{(3)}$ and $\Delta A_{gg,Q}^{(3)}$ \cite{Ablinger:2022wbb} this 
turned out not to be the 
case down to the lowest moment for a single diagram, for which $x$-space technologies
had to be used to obtain the correct physical expression. The evanescent sum-product 
solution valid 
for all higher moments thus turns out to be a special case of a global higher 
transcendental 
solution. Techniques implemented
in the package {\tt HarmonicSums.m} 
\cite{Ablinger:2013eba,Ablinger:2014rba,Ablinger:2010kw,Ablinger:2013hcp,ALL2016,
Ablinger:2015gdg,Ablinger:2018cja,Ablinger:2019mkx,ALL2018,
Blumlein:2009ta,Vermaseren:1998uu,Blumlein:1998if,
Remiddi:1999ew,Ablinger:2011te,Ablinger:2013cf,Ablinger:2014bra,Blumlein:2003gb,
Blumlein:2009cf,Ablinger:2013jta}  
allow to derive the associated $x$-space 
representations and offers various methods to simplify the results.

In the case of the single- and two-mass OMEs $A_{Qg}^{(3)}$ and $\Delta A_{Qg}^{(3)}$ these 
representations are not sufficient. Although one may obtain recurrences for these quantities
in $N$-space, they are not first-order-factorizable. Also the differential equations
in $x$-space do not 
factorize to first order. The corresponding iterated integrals
contain higher transcendental letters. They can be calculated using the 
method described in  Ref.~\cite{Behring:2023rlq}. The local operators defined for discrete integer 
values of $N$
are resummed into a generating function which depends on the continuous real variable $t 
\in [0,\infty[$,
cf.~Refs.~\cite{Ablinger:2012qm,Ablinger:2014yaa}. Furthermore, before treating the 
regular part, one has to subtract the distribution-valued contributions.

The master integrals are computed in this
variable by solving linear systems of coupled differential equations,
see also Refs.~\cite{Ablinger:2018zwz,Maier:2017ypu,Fael:2021kyg,Fael:2022miw,Ablinger:2023ahe}. 
The initial values are provided by the 
Mellin moments, which are the expansion coefficients at $t=0$ or $x \rightarrow \infty$.
One solves the differential equations in $t$ and performs then the analytic continuation
from the region $t \in [0,1]$ to $t \in [1,\infty[$.
For $t < 1$ the amplitude does not contain an imaginary part. Still one 
has to
perform a series of matchings of local expansions until one reaches the point $t=1$. 
The variables $t$ and $x$ are related via $x = 1/t$.
In the region $x \in [0,1]$ the solutions are expanded into different overlapping 
logarithmic modulated Taylor series. This representation also provides the numerical results.  

In the two-mass cases the gluonic OMEs can be represented by iterated integrals
over root-valued letters, see Refs.~\cite{Ablinger:2018brx,Ablinger:2020snj}. 
Also many new special constants beyond the multiple zeta values 
\cite{Blumlein:2009cf} are contributing. In the pure-singlet case, only the $x$-space 
representations 
obey first-order-factorizing differential equations \cite{Ablinger:2017xml, 
Ablinger:2019gpu}, which 
have to be solved for sub-intervals in $x \in [0,1]$. Finally, the two-mass corrections for 
$A_{Qg}^{(3)}$ and $\Delta A_{Qg}^{(3)}$ \cite{Ablinger:2025nnq} are calculated in a similar way as 
the ones in the single-mass case. The best convergence is obtained by expanding in the 
mass ratio 
around $m_c = m_b$ and by additional use of convergence acceleration due to Aitken extrapolation 
\cite{AITKEN}.

There is a variety of other calculation methods which have been developed and used in 
this 
project. A survey on these computer-algebraic and  mathematical techniques are given in 
Refs.~\cite{Blumlein:2018cms,Blumlein:2021ynm}.
\section{The single-mass corrections} 
\label{sec:3} 

\vspace*{1mm}
\noindent
The extrinsic inclusive heavy-quark corrections to deep-inelastic scattering in the 
unpolarized and 
polarized case were obtained in Refs.~\cite{Witten:1975bh,Babcock:1977fi,Shifman:1977yb,
Leveille:1978px} and \cite{Watson:1981ce}. At two-loop order, numerical results for the 
complete
kinematics in the unpolarized case were obtained in 
Refs.~\cite{Laenen:1992zk,Laenen:1992xs,Riemersma:1994hv}, and in the
polarized case in Ref.~\cite{Hekhorn:2018ywm}. Mellin-space implementations were given in 
Ref.~\cite{Alekhin:2003ev}. Analytic two-loop calculations were performed 
in the unpolarized and 
polarized non-singlet cases in Refs.~\cite{Buza:1995ie,Buza:1996xr,Blumlein:2016xcy} and 
in the pure-singlet cases in Refs.~\cite{Blumlein:2019qze,Blumlein:2019zux}.

Analytic two-loop calculations of the heavy-flavor Wilson coefficients in the region $Q^2 \gg m_Q^2$ 
were independently performed in Refs.~\cite{Buza:1995ie,Bierenbaum:2007qe} in the unpolarized case 
and Refs.~\cite{Buza:1996xr,Bierenbaum:2022biv} in  the polarized case.
Further massive OMEs contributing to the 
variable flavor number scheme  were calculated to two-loop orders in 
Refs.~\cite{Buza:1996wv,Bierenbaum:2009zt}.

The logarithmic three-loop corrections are based on the complete two-loop results for the massive 
OMEs and the three-loop massless Wilson coefficients 
\cite{Vermaseren:2005qc,Moch:2008fj,Blumlein:2022gpp}. 
These contributions were given in Refs.~\cite{Kawamura:2012cr,Behring:2014eya} in the 
unpolarized case and in Ref.~\cite{Blumlein:2021xlc} in the polarized case.

Also the asymptotic three-loop corrections to the structure function $F_L(x,Q^2)$ depend only on the 
two-loop massive OMEs and were calculated in Ref.~\cite{Blumlein:2006mh}. It turns out 
that these 
corrections, already at NLO, describe the full phase-space results only for $Q^2/m_Q^2 > 800~\GeV^2$,
i.e. in a kinematic region, where there are currently no data.

The asymptotic massive three-loop Wilson coefficients for the unpolarized structure function 
$F_2(x,Q^2)$ and polarized structure function $g_1(x,Q^2)$ were calculated in 
Refs.~\cite{Behring:2014eya,Ablinger:2014vwa,Ablinger:2014nga,Ablinger:2023ahe,
Ablinger:2024xtt,Ablinger:2025awb}.
and Refs.~\cite{Ablinger:2014vwa,Behring:2015zaa,
Blumlein:2021xlc,  
Ablinger:2023ahe,Ablinger:2019etw,
Ablinger:2024xtt,Ablinger:2025awb}. Where necessary, we used the Larin scheme. Here, also 
the parton
distributions have to be evolved in this scheme, see Ref.~\cite{Blumlein:2024euz} for a 
parameterization.

Also the heavy-flavor corrections to deep-inelastic charged current structure functions 
were 
calculated. Here, also flavor excitation processes, e.g. for $s \rightarrow c$, 
contribute.
For the one-loop order cross sections see Refs.~\cite{Gluck:1996ve,Blumlein:2011zu}.
At two-loop order, there are results in the asymptotic region 
\cite{Buza:1997mg,Blumlein:2014fqa}.
Furthermore, there are three-loop heavy-flavor corrections to the non-singlet structure 
functions
$xF_3(x,Q^2)$ \cite{Behring:2015roa} and the differences $F_L^{W^+-W^-}(x,Q^2)$ and 
$F_2^{W^+-W^-}(x,Q^2)$ \cite{Behring:2016hpa}.

The massive OMEs also determine the flavor matching in the variable flavor number scheme (VFNS).
The two-loop single-mass matching relations were given in Ref.~\cite{Buza:1996wv} with
corrections given in Ref.~\cite{Bierenbaum:2009zt} in the unpolarized case. The further two-loop 
massive OMEs in the polarized case were calculated in Ref.~\cite{Bierenbaum:2022biv}. At three-loop 
order the VFNS was studied in Ref.~\cite{Ablinger:2025joi}.\footnote{For the use in 
evolution codes see e.g. Ref.~\cite{Karlberg:2025hxk}.}
 The additional OMEs $A_{gq}$ and $A_{gg}$
were calculated in Refs.~\cite{Ablinger:2014lka,Ablinger:2022wbb}  in the unpolarized case and 
Refs.~\cite{Behring:2021asx,Ablinger:2022wbb} in the 
polarized case.  

Finally, also the massive three-loop corrections have been calculated to the massive OME for 
transversity in Refs.~\cite{Blumlein:2009rg,Ablinger:2014vwa}.
\begin{figure}[h]
\centering
\includegraphics[width=0.49\textwidth]{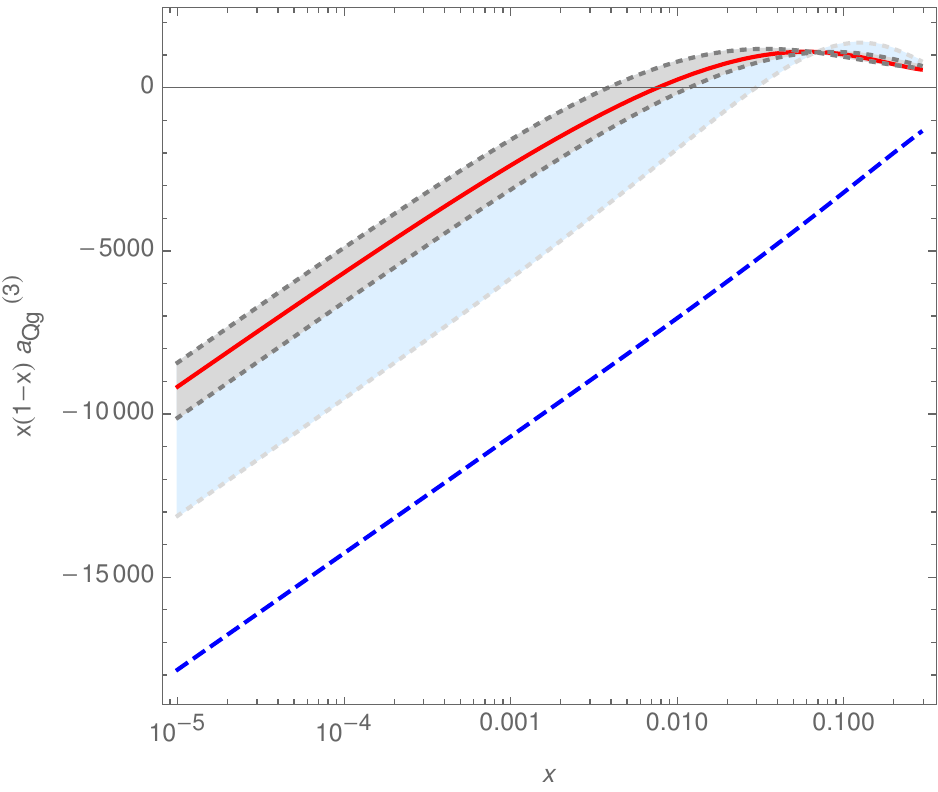}
\includegraphics[width=0.49\textwidth]{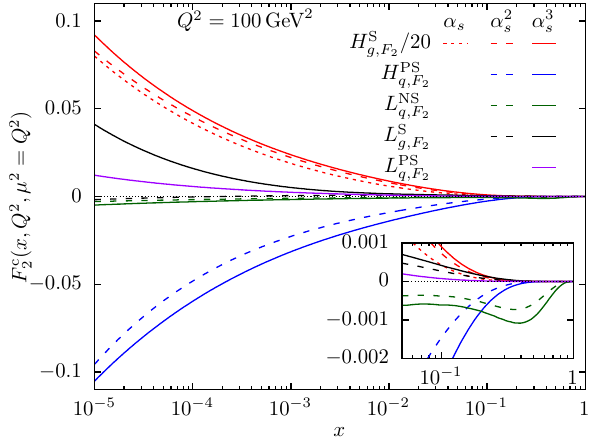}
\caption{\sf Left panel: 
the constant part of the massive OME $A_{Qg}^{(3)}$, $a_{Qg}^{(3)}(x)$,  
as a
function of $x$, rescaled by the factor $x(1-x)$ in the region of
smaller values of $x$.
Full line (red): $a_{Qg}^{(3)}(x)$;
dashed line (blue): leading small-$x$ term $\propto \ln(x)/x$
\cite{Catani:1990eg};
light blue region: estimates of \cite{Kawamura:2012cr};
gray region: estimates of \cite{Alekhin:2017kpj}; from Ref.~\cite{Ablinger:2024xtt}.
Right panel: the charm contributions to the structure function
$F_2(x,Q^2)$ by the Wilson coefficients  $H_g^{\rm S}, H_q^{\rm PS}, L_q^{\rm NS}, 
L_g^{\rm
S}$ and
$L_g^{\rm PS}$ at $Q^2 = 100~\GeV^2$ at different orders in the strong coupling constant
up to $O(a_s), O(a_s^2)$ and $O(a_s^3)$; from Ref.~\cite{Ablinger:2025awb}. For the 
notations of the massive Wilson coefficients see Ref.~\cite{Bierenbaum:2009mv}.}
\label{fig:1}
\end{figure}

\noindent
In Figure~\ref{fig:1}, we illustrate the constant part of the unpolarized unrenormalized 
three-loop single-mass OME $A_{Qg}$, $a_{Qg}^{(3)}(x)$, 
Ref.~\cite{Ablinger:2023ahe,Ablinger:2024xtt}. 
Previous approximate estimates with error bands are now replaced by the exact result.
We also show the result for the leading small-$x$ term $\propto \ln(x)/x$, which does not describe 
the result. As also in various other cases \cite{Blumlein:1997em,Blumlein:1998mg,Blumlein:1999ev}
even in the small-$x$ region a series of 
sub-leading terms are of the same size as the leading term or even larger and cause this 
difference.
In an earlier analysis, there was a theory error of $\delta_T m_c = 70$~MeV, much larger 
than the 
experimental 
error \cite{Alekhin:2012vu}. This error can now be significantly reduced.

We also illustrate the different contributions to the heavy-flavor corrections, e.g. for 
the charm 
contributions in Figure~\ref{fig:1}. The positive gluonic terms are largest, followed by the
negative pure-singlet terms in the small-$x$ region. There are three other contributions 
which are 
smaller but are relevant at resolutions of $O(1\%)$. In the large-$x$ region the non-singlet 
contribution dominates and is negative. That means that there the massless contributions 
to $F_2(x,Q^2)$ are 
diminished by the heavy-flavor contributions. The structure function is still positive.   

\section{The two-mass corrections} 
\label{sec:4} 

\vspace*{1mm}
\noindent
Two-mass heavy-flavor corrections arise first at two-loop order as factorizable vacuum 
polarization 
contributions in one-loop graphs for the OMEs $A_{Qg}$ and $A_{gg,Q}$, 
cf.~Ref.~\cite{Blumlein:2018jfm}, in the unpolarized case and Ref.~\cite{Bierenbaum:2022biv} in the
polarized case. Genuine two-mass corrections arise from three-loop order and contribute to the 
OMEs $A_{qq,Q}^{\rm NS}, A_{qq,Q}^{\rm NS, trans}$ and $A_{gq,Q}$, calculated in 
Ref.~\cite{Ablinger:2017err}. 
The unpolarized and polarized two-mass OMEs $A_{Qq}^{\rm PS}$ were calculated in
Refs.~\cite{Ablinger:2017xml,Ablinger:2020snj}, the ones for $A_{gg,Q}$ in 
Refs.~\cite{Ablinger:2018brx, Ablinger:2020snj}, for $\Delta A_{gq,Q}$ in 
Ref.~\cite{Behring:2021asx}. Finally, the two-mass corrections to
$A_{Qg}$ in the unpolarized and polarized case were computed in 
Ref.~\cite{Ablinger:2025nnq}.

\begin{figure}[h]\centering
\includegraphics[width=0.5\textwidth]{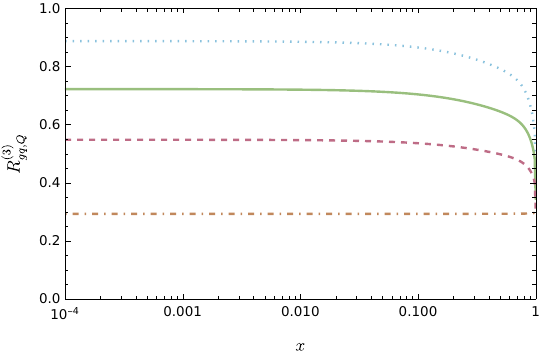}
\caption[]{\sf
The ratio of the three-loop two-mass contributions to $A_{gq,Q}$ to the complete {gq} 
$O(T_F^2)$ corrections. 
Dotted line: $Q^2 = 30~\GeV^2$; 
Full line: $Q^2 = 50~\GeV^2$; 
Dashed line: $Q^2 = 100~\GeV^2$; 
Dashed-dotted: $Q^2 = 1000~\GeV^2$.
}
\label{fig:2}
\end{figure}

Because the charm and bottom mass are very similar, one may consider to decouple both contributions
together in a two-mass VFNS, cf.~\cite{Blumlein:2018jfm,Ablinger:2017err}.
The two-mass corrections to the structure functions emerge as $O(T_F^2 C_{F,A})$ 
contributions.
We computed the relative two-mass part in all $T_F^2$ contributions. On average it 
amounts to $O(50\%)$ for all processes. We 
illustrate this for the two-mass contributions to $A_{gq}^{(3)}$ in Figure~\ref{fig:2}. 

\section{The structure functions} 
\label{sec:5} 

\vspace*{1mm}
\noindent
\begin{figure}[h]\centering
\includegraphics[width=0.48\textwidth]{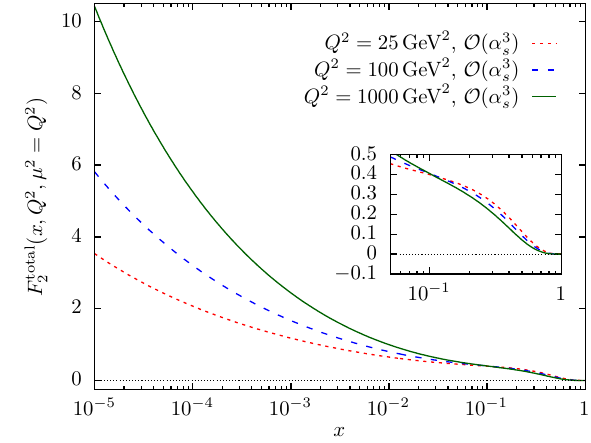}
\includegraphics[width=0.48\textwidth]{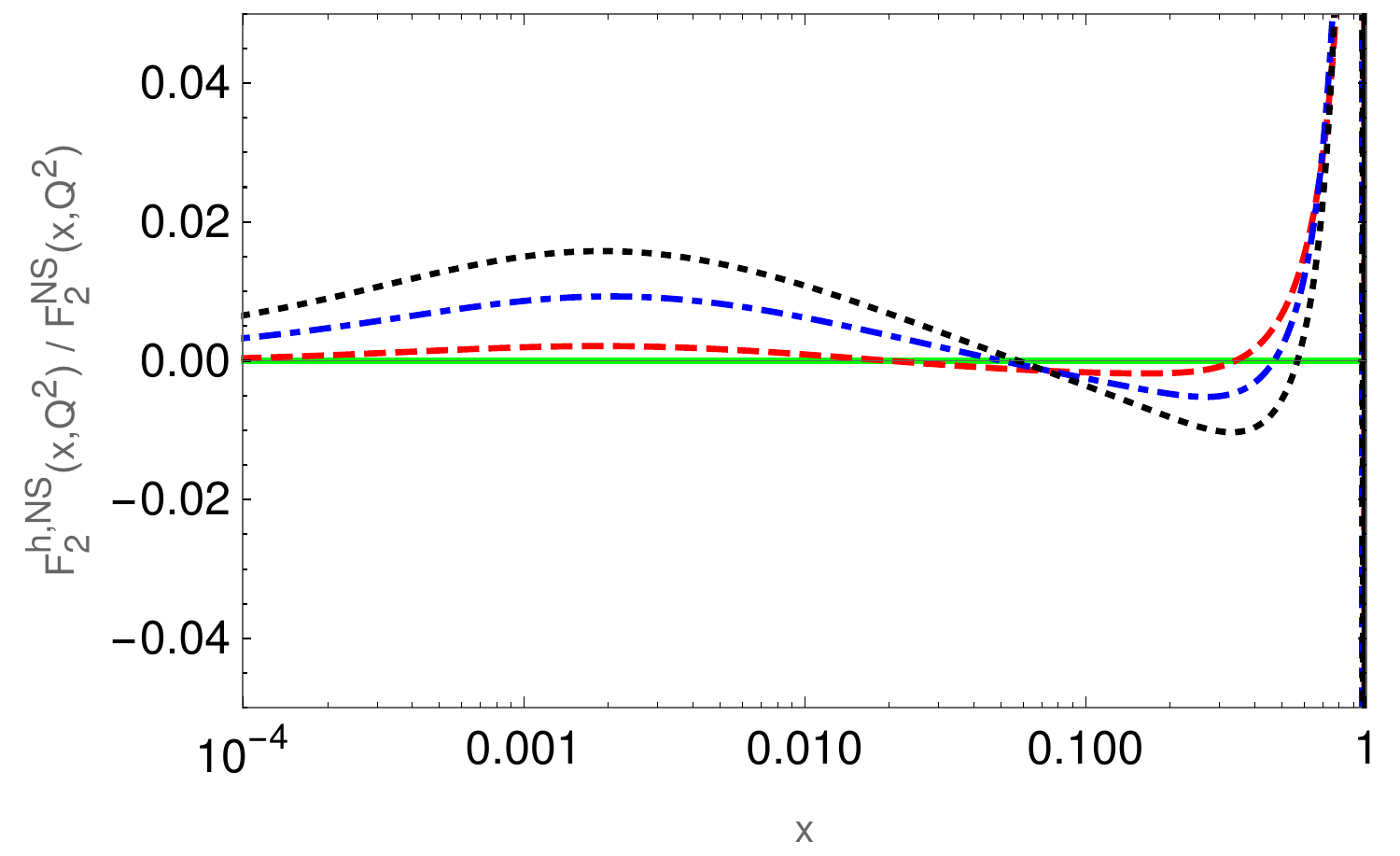}
\caption[]{\sf Left panel: The structure function $F_2(x,Q^2)$ at NNLO containing the 
massless and 
the asymptotic single-mass contributions due to charm and bottom quarks as a function 
of $x$ and 
$Q^2$, using the parton distribution functions of Ref.~\cite{Alekhin:2017kpj}; from 
Ref.~\cite{Ablinger:2025awb}.
Right panel: 
The relative contribution of the heavy-flavor contributions due
to $c$ and $b$ quarks
to the structure function $F_2^{\rm NS}$; dashed lines:
$100~\GeV^2$;  dashed-dotted lines: $1000~\GeV^2$;
dotted lines: $10000~\GeV^2$. massless: N$^3$LO contributions and single- and two-mass 
contributions: $O(a_s^3)$; from Ref.~\cite{Blumlein:2021lmf}.
}
\label{fig:3}
\end{figure}

\noindent 
The present results on the asymptotic heavy-flavor Wilson coefficients allow the NNLO 
QCD-analysis of the deep-inelastic structure function data on $F_2(x,Q^2)$ at large enough 
virtualities for the first time. Corresponding results were also obtained for the 
polarized structure function $g_1(x,Q^2)$.
Previous analyses were still lacking parts of the 
heavy-flavor corrections.
Since the heavy-flavor corrections have different scaling violations than the massless
ones, these corrections are essential.
In Figure~\ref{fig:3}, we show the prediction of the scale evolution of $F_2(x,Q^2)$ in a 
wide range of $x$ and $Q^2$, referring to the parton distribution functions of 
Ref.~\cite{Alekhin:2017kpj}.

As well-known, the measured values of $\alpha_s(M_Z)$ from deep-inelastic scattering data 
of different analyses do not yet agree. In the future there will be dedicated measurements 
at the EIC \cite{Boer:2011fh}, using both proton and deuteron data.
 From those, 
by performing
deuteron wave-function corrections, one can extract a flavor non-singlet structure function.
For this quantity one may study scheme-invariant scale evolution, by measuring the input
at a scale $Q_0^2$ and parameterizing it to sufficient precision. The scale evolution 
is then driven by the QCD-scale $\Lambda_{\rm QCD}$ only, still being correlated to the 
measured input. These data are not affected by the gluon distribution and various sea-quark 
distributions. Here, the heavy-flavor corrections are rather small and predictions have 
been 
made at N$^3$LO in Ref.~\cite{Blumlein:2021lmf} already, see Figure~\ref{fig:3}.\footnote{
The study of electron-deuteron scattering has also been considered \cite{Blumlein:1987xk}
and 
later proposed for HERA \cite{DESYPRC} but was not realized.}
We regard this as the cleanest way
to extract $\alpha_s(M_Z^2)$ from deep-inelastic scattering data, through which the present
debates can be settled.\footnote{A measurement of this kind had been proposed for an
upgrade of the BCDMS experiment in 1989 by C.~Guyot et al. in Ref.~\cite{STAUDE} amounting 
to 35 million CHF. This proposal was not realized. The proposed experiment could have 
answered the still 
open 
question on the value of $\alpha_s(M_Z^2)$ from deep-inelastic data.}
It is needless to say, that also here the region of higher 
twist corrections has to be cut out in the analysis.

\section{Numerical Implementations of Massless Wilson Coefficients, Splitting 
Functions and Target-Mass Corrections}
\label{sec:6}

\vspace*{1mm}
\noindent
We provide the {\tt Fortran} libraries {\tt WILS3}, {\tt SPLIT\_U}, and  {\tt SPLIT\_P} 
for the massless Wilson coefficients of deep-inelastic scattering and the unpolarized and 
polarized splitting functions to three-loop order as fast and precise numerical 
implementations, in a similar form as for the implementation of the massive asymptotic 
Wilson 
coefficients and massive operator matrix elements as given in 
Refs.~\cite{Ablinger:2025awb,Ablinger:2025joi}. For the polarized splitting functions we
provide both the splitting functions in the M-scheme \cite{Matiounine:1998re} and the 
Larin scheme 
\cite{Larin:1993tq}. We also include the code {\tt TARGM} for the target-mass corrections
for the structure functions $F_2, F_L, xF_3, g_1$ and $g_2$, see 
Ref.~\cite{Georgi:1976ve}\footnote{Here we 
correct the known typographical errors in the expressions for the structure functions $F_1$ 
and $xF_3$, see also Ref.~\cite{Schienbein:2007gr}.} 
and Refs.~\cite{Piccione:1997zh,Blumlein:1998nv}. 
They are well suited to be used in $x$-space evolution codes for the QCD analysis of 
deep-inelastic scattering data.\footnote{As target-mass corrections are 
structure-function related, one may consider to correct the data for these effects prior
to QCD-fits, as has been done Ref.~\cite{Christy:2012tk}. Because of the 
emerging double-integrals,
this will save time in the QCD-fitting codes. Likewise, one also performs QED radiative
corrections \cite{Kwiatkowski:1990es,Arbuzov:1995id} on the data before analyzing the 
scaling violations.} 

The Wilson coefficients are represented as polynomials
in the parameters 
\begin{equation}
\{{\tt IO}, a_s, x, N_F, \ln(Q^2/\mu^2), {\tt flav}\}, \nonumber 
\end{equation}
see 
Ref.~\cite{Ablinger:2025awb}. 
The $x$-dependence is described by elementary functions and polynomial interpolations.
The order in the coupling constant is selected by the parameter {\tt IO} = 1,2,3.
We include the Wilson coefficients for the structure functions $F_2(x,Q^2), F_L(x,Q^2),
xF_3(x,Q^2)$ \cite{Vermaseren:2005qc,Moch:2008fj,Blumlein:2022gpp}
and $g_1(x,Q^2)$ \cite{Blumlein:2022gpp}. For the splitting functions, we 
refer to Refs.~\cite{Moch:2004pa,Ablinger:2014nga,Ablinger:2017tan,Blumlein:2021enk,
HADR,Duhr:2020seh,Gehrmann:2023ksf} 
in the 
unpolarized case and 
Refs.~\cite{Moch:2014sna,Blumlein:2021ryt,Behring:2025avs,Zhu:2025gts} in the 
polarized case. The representations are based on analytic 
expansions in the small and large-$x$ regions and a numerical representation of the 
remainder part, see Ref.~\cite{Ablinger:2025awb} for details. Unlike the case in Mellin-$N$
space, one has to split the parts $\propto \delta(1-x)$ {\tt DEL...f}, the +-distribution 
contributions {\tt PLU...f}, and the 
regular parts into different functions in $x$-space, because of the different 
Mellin-convolutions
in $x$-space, cf.~Eqs.~(12--14) of Ref.~\cite{Ablinger:2025awb}. The regular parts are given by a 
superposition of the small-$x$ {\tt SX...f}, large-x {\tt LX...f}, and a remainder contribution
{\tt GR...f},  based on cubic spline-interpolation \cite{SPLINE} of fine grids for the individual 
polynomial contributions. The $\delta(1-x)$- and +-function terms, as well as the small and large-$x$ 
expansions are coded as analytic expressions, for which we used code-optimization 
\cite{Ruijl:2017dtg}. In the case of the polarized splitting functions, 
the above argument list is extended by the parameter {\tt LARIN} = 0 (M-scheme), = 1 
(Larin scheme).
The parameterization of the splitting functions is given with a factor of 1/2 relative to
the ones in Refs.~\cite{Blumlein:2021enk,Blumlein:2021ryt}.
For the target-mass corrections, we also include 
the numerical integration routines {\tt DAIND} of Ref.~\cite{AIND}.

The license conditions for any use of these codes 
require the citation of this paper and the references on which their implementation have 
been based. They are given in separate files along with the other parts of the codes.

\section{Conclusions}
\label{sec:7}

\vspace*{1mm}
\noindent
Given the current precision of the World deep-inelastic data and facing high luminosity data 
taken at the EIC in the near future, the theoretical description of the scale evolution
of the deep-inelastic structure function has to be more precise than the $O(1\%)$-level.
Here one missing asset has been the three-loop heavy-flavor corrections, now being 
available in the 
region of large $Q^2$. Future dedicated precision measurements of $\alpha_s(M_Z^2)$ will  be 
performed under these conditions to unambiguously determine this fundamental coupling constant.

Along with this calculation project, various new computer-algebraic and mathematical 
algorithms,
tools and techniques were developed, which are of wider use in other perturbative higher order
calculations in QED, QCD, and the Standard Model. 

The results of the project allow the analysis of both unpolarized and polarized precision data and
will help to improve the accuracy of the extraction of $m_c$ and $m_b$ using upcoming high
luminosity data. With this new level of precision also more precise parton distributions will be 
obtained and a deeper insight into the nucleon spin problem is possible.
The presently available theoretical framework will improve re-analyses of the current deep-inelastic 
data and play a central role in analyzing upcoming data at the EIC and for proposed projects like 
the LHeC \cite{LHeCStudyGroup:2012zhm,LHeC:2020van}.

We also provide fast and precise numerical implementations of the unpolarized and polarized 
three-loop massless Wilson coefficients and splitting functions in terms of public codes.

\vspace*{5mm}
\noindent
{\bf Acknowledgments.}
We thank P.~Marquard for
discussions. This work was supported by the European Research Council (ERC) 
under the European Union's Horizon 2020 research and innovation programme   
grant agreement 101019620 (ERC Advanced Grant TOPUP), the UZH Postdoc Grant,
grant no.~[FK-24-115] and by the Austrian Science Fund (FWF) Grant-DOI 10.55776/P20347.
KS is support by the European Union under the HORIZON program in Marie Sklodowska-Curie
project No. 101204018.
\hspace*{-2cm} 
\parbox{180pt}{\centering\includegraphics[height=5mm]{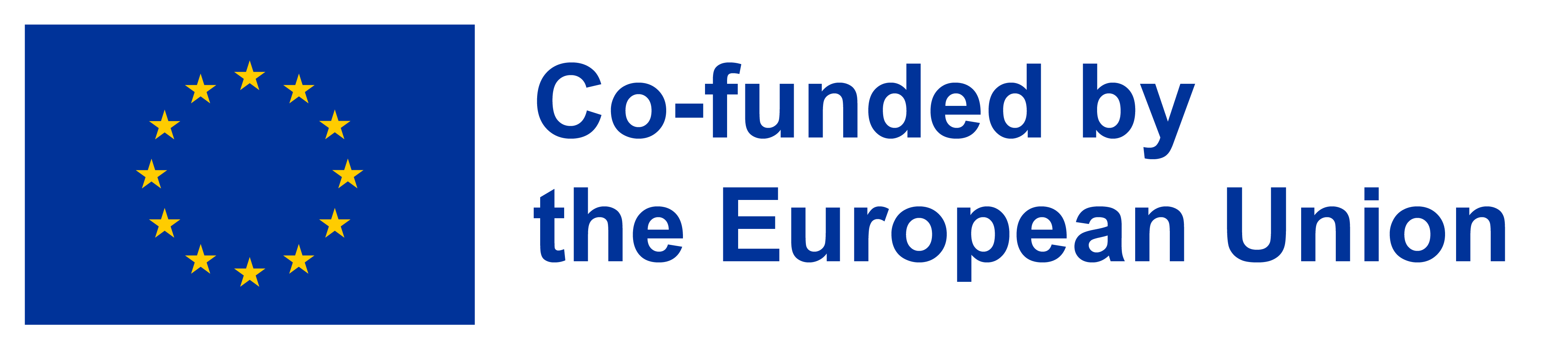}.}

{\footnotesize

}

\begin{thebibliography}{99}
%
\bibitem{Bethke:2011tr}
  S.~Bethke {\it et al.},
  {\it Workshop on Precision Measurements of $\alpha_s$},
  arXiv:1110.0016 [hep-ph].
%
\bibitem{Moch:2014tta}
  S.~Moch {\it et al.},
  {\it High precision fundamental constants at the TeV scale},
  arXiv:1405.4781 [hep-ph].
%
\bibitem{Alekhin:2016evh}
  S.~Alekhin, J.~Bl\"umlein and S.O.~Moch,
  Mod.\ Phys.\ Lett.\ A {\bf 31} (2016) no.25,  1630023.
%
\bibitem{dEnterria:2022hzv}
D.~d'Enterria \textit{et al.}, 
J. Phys. G \textbf{51} (2024) no.9, 090501
[arXiv:2203.08271 [hep-ph]].
%
\bibitem{Alekhin:2012vu}
S.~Alekhin, J.~Bl\"umlein, K.~Daum, K.~Lipka and S.~Moch,
Phys. Lett. B \textbf{720} (2013) 172--176
[arXiv:1212.2355 [hep-ph]].
%
\bibitem{Blumlein:2006be}
J.~Bl\"umlein, H.~B\"ottcher and A.~Guffanti,
Nucl. Phys. B \textbf{774} (2007) 182--207
[hep-ph/0607200].
%
\bibitem{Blumlein:2008kz}
J.~Bl\"umlein and H.~B\"ottcher,
Phys. Lett. B \textbf{662} (2008) 336--340
[arXiv:0802.0408 [hep-ph]].
%
\bibitem{Blumlein:2012se}
J.~Bl\"umlein and H.~B\"ottcher,
Proc. of DIS 2012, Bonn, 2012, 237--241, DESY-PROC-2012-02,
[arXiv:1207.3170 [hep-ph]].
%
\bibitem{Alekhin:2012ig}
S.~Alekhin, J.~Bl\"umlein and S.~Moch,
Phys. Rev. D \textbf{86} (2012) 054009
[arXiv: 1202.2281 [hep-ph]].
%
\bibitem{Buza:1995ie}
  M.~Buza, Y.~Matiounine, J.~Smith, R.~Migneron and W.L.~van Neerven,
  Nucl.\ Phys.\ B {\bf 472} (1996) 611--658
  [hep-ph/9601302].
%
\bibitem{Laenen:1992zk}
E.~Laenen, S.~Riemersma, J.~Smith and W.L.~van Neerven,
Nucl. Phys. B \textbf{392} (1993) 162--228.
%
\bibitem{Laenen:1992xs}
E.~Laenen, S.~Riemersma, J.~Smith and W.L.~van Neerven,
Nucl. Phys. B \textbf{392} (1993) 229--250.
%
\bibitem{Riemersma:1994hv}
S.~Riemersma, J.~Smith and W.L.~van Neerven,
Phys. Lett. B \textbf{347} (1995) 143--151
[hep-ph/9411431].
%
\bibitem{Vermaseren:1998uu}
  J.A.M.~Vermaseren,
  { Int.\ J.\ Mod.\ Phys.}\ A {\bf 14} (1999) 2037--2076
  [hep-ph/9806280].
%
\bibitem{Blumlein:1998if}
  J.~Bl\"umlein and S.~Kurth,
  {Phys.\ Rev.}\ D {\bf 60} (1999) 014018
  [hep-ph/9810241].
%
\bibitem{Moch:2001zr}
S.~Moch, P.~Uwer and S.~Weinzierl,
J. Math. Phys. \textbf{43} (2002) 3363--3386
[hep-ph/0110083].
%
\bibitem{Ablinger:2013cf}
  J.~Ablinger, J.~Bl\"umlein and C.~Schneider,
  J.\ Math.\ Phys.\  {\bf 54} (2013) 082301
  [arXiv:1302.0378 [math-ph]].  
%
\bibitem{Ablinger:2011te}
J.~Ablinger, J.~Blumlein and C.~Schneider,
J. Math. Phys. \textbf{52} (2011), 102301
[arXiv:1105.6063 [math-ph]].
%
\bibitem{Ablinger:2014bra}
  J.~Ablinger, J.~Bl\"umlein, C.G.~Raab and C.~Schneider,
  J.\ Math.\ Phys.\  {\bf 55} (2014) 112301
  [arXiv:1407.1822 [hep-th]].   
%
\bibitem{Ablinger:2021fnc}
J.~Ablinger, J.~Bl{\"u}mlein and C.~Schneider,
Phys. Rev. D \textbf{103} (2021) no.9, 096025
[arXiv:2103.08330 [hep-th]].
%
\bibitem{Ablinger:2017bjx}
J.~Ablinger, J.~Bl\"umlein, A.~De Freitas, M.~van Hoeij, E.~Imamoglu,
C.G.~Raab, C.S.~Radu and C.~Schneider,
J. Math. Phys. \textbf{59} (2018)  062305
[arXiv:1706.01299 [hep-th]].
%
\bibitem{JB2016}
J.~Bl\"umlein, talk
The 5th International Congress on Mathematical Software ZIB Berlin from July 11
to July 14, 2016, Session: Symbolic computation and elementary particle physics,\\
{\tt https://www.risc.jku.at/conferences/ICMS2016/};
see also J.B.'s talk at 
QCD@LHC, U. Z\"u- rich, 22--26.08. 2016,
{\tt https://indico.cern.ch/event/516210/timetable/\#all.detailed}.
%
\bibitem{Remiddi:1999ew}
E.~Remiddi and J.A.M.~Vermaseren,
Int. J. Mod. Phys. A \textbf{15} (2000) 725--754
[hep-ph/9905237].
%
\bibitem{Blumlein:2003gb}
J.~Bl\"umlein,
Comput. Phys. Commun. \textbf{159} (2004) 19--54
[hep-ph/0311046].
%
\bibitem{Gehrmann:2001pz}
T.~Gehrmann and E.~Remiddi,
  Comput.\ Phys.\ Commun.\  {\bf 141} (2001) 296--312
  [hep-ph/0107173].
%
\bibitem{Ablinger:2018sat}
J.~Ablinger, J.~Bl\"umlein, M.~Round and C.~Schneider,
Comput. Phys. Commun. \textbf{240} (2019) 189--201
[arXiv:1809.07084 [hep-ph]].
%
\bibitem{Vollinga:2004sn}
J.~Vollinga and S.~Weinzierl,
Comput. Phys. Commun. \textbf{167} (2005) 177--194
[hep-ph/0410259].
%
\bibitem{Borwein:1999js}
J.M.~Borwein, D.M.~Bradley, D.~J.Broadhurst and P.~Lisonek,
Trans. Am. Math. Soc. \textbf{353} (2001) 907--941
[arXiv:math/9910045 [math.CA]].
%
\bibitem{Behring:2023rlq}
A.~Behring, J.~Bl{\"u}mlein and K.~Sch{\"o}nwald,
JHEP \textbf{06} (2023) 062
[arXiv:2303.05943 [hep-ph]].
%
\bibitem{Buza:1996xr}
M.~Buza, Y.~Matiounine, J.~Smith and W.~L.~van Neerven,
Nucl. Phys. B \textbf{485} (1997) 420--456
[hep-ph/9608342].
%
\bibitem{Blumlein:2016xcy}
J.~Bl{\"u}mlein, G.~Falcioni and A.~De Freitas,
Nucl. Phys. B \textbf{910} (2016)  568--617
[arXiv:1605.05541 [hep-ph]].
%
\bibitem{Blumlein:2019qze}
J.~Bl\"umlein, A.~De Freitas, C.G.~Raab and K.~Sch\"onwald,
Nucl. Phys. B \textbf{945} (2019) 114659
[arXiv:1903.06155 [hep-ph]].
%
\bibitem{Blumlein:2019zux}
J.~Bl{\"u}mlein, C.~Raab and K.~Sch{\"o}nwald,
Nucl. Phys. B \textbf{948} (2019) 114736
[arXiv:1904.08911 [hep-ph]].
%
\bibitem{Alekhin:2003ev}
S.I.~Alekhin and J.~Bl\"umlein,
Phys. Lett. B \textbf{594} (2004) 299--307
[hep-ph/0404034].
%
\bibitem{Ablinger:2025joi}
J.~Ablinger, A.~Behring, J.~Bl{\"u}mlein, A.~De Freitas, A.~von Manteuffel, C.~Schneider 
and K.~Sch{\"o}nwald,
{\it The Single-Mass Variable Flavor Number Scheme at Three-Loop Order},
[arXiv:2510.02175 [hep-ph]], JHEP (2026) in print.
%
\bibitem{Ablinger:2025awb}
J.~Ablinger, A.~Behring, J.~Bl{\"u}mlein, A.~De Freitas, A.~von Manteuffel, C.~Schneider 
and K.~Sch{\"o}nwald,
{\it The three-loop single-mass heavy-flavor corrections to the structure functions 
$F_2(x,Q^2)$ and $g_1(x,Q^2)$},
[arXiv:2509.16124 [hep-ph]].
%
\bibitem{YND}
F.~Yndurain, {\sf The Theory of Quark and Gluon Interactions}, (Springer, Berlin 2006).
%
\bibitem{Bierenbaum:2009mv}
  I.~Bierenbaum, J.~Bl\"umlein and S.~Klein,
  Nucl.\ Phys.\ B {\bf 820} (2009) 417--482
  [arXiv:0904.3563 [hep-ph]].
%
\bibitem{Ablinger:2025nnq}
J.~Ablinger, J.~Bl{\"u}mlein, A.~De Freitas, A.~von Manteuffel, C.~Schneider and K.~Sch{\"o}nwald,
JHEP \textbf{01} (2026) 111
[arXiv:2510.09403 [hep-ph]]
%
\bibitem{Ablinger:2017err}
  J.~Ablinger, J.~Bl\"umlein, A.~De Freitas, A.~Hasselhuhn, C.~Schneider and F.~Wi\ss{}brock,
  Nucl.\ Phys.\ B {\bf 921} (2017) 585--688
  [arXiv:1705.07030 [hep-ph]].
%
\bibitem{Buza:1996wv}
  M.~Buza, Y.~Matiounine, J.~Smith and W.L.~van Neerven,
  Eur.\ Phys.\ J.\ C {\bf 1} (1998) 301--320
  [hep-ph/9612398].
%
\bibitem{Vermaseren:2005qc}
J.A.M.~Vermaseren, A.~Vogt and S.~Moch,
Nucl. Phys. B \textbf{724} (2005) 3--182
[hep-ph/0504242].
%
\bibitem{Moch:2008fj}
S.~Moch, J.A.M.~Vermaseren and A.~Vogt,
Nucl. Phys. B \textbf{813} (2009) 220--258
[arXiv:0812.4168 [hep-ph]].
%
\bibitem{Blumlein:2022gpp}
J.~Bl\"umlein, P.~Marquard, C.~Schneider and K.~Sch\"onwald,
JHEP \textbf{11} (2022) 156
[arXiv:2208.14325 [hep-ph]].
%
\bibitem{Ablinger:2023ahe}
J.~Ablinger, A.~Behring, J.~Bl\"umlein, A.~De Freitas, A.~von Manteuffel, 
C.~Schneider and K.~Sch\"onwald,
Nucl. Phys. B \textbf{999} (2024) 116427
[arXiv:2311.00644 [hep-ph]].
%
\bibitem{Nogueira:1991ex}
  P.~Nogueira,
  J.\ Comput.\ Phys.\  {\bf 105} (1993) 279--289.
%
\bibitem{Vermaseren:2000nd}
  J.A.M.~Vermaseren,
  {\it New features of FORM},
  math-ph/0010025.
%
\bibitem{Tentyukov:2007mu}
  M.~Tentyukov and J.A.M.~Vermaseren,
  Comput.\ Phys.\ Commun.\  {\bf 181} (2010) 1419--1427
  [hep-ph/0702279].
%
\bibitem{vanRitbergen:1998pn}
T.~van Ritbergen, A.N.~Schellekens and J.A.M.~Vermaseren,
Int. J. Mod. Phys. A \textbf{14} (1999) 41--96
[hep-ph/9802376].
%
\bibitem{Studerus:2009ye}
  C.~Studerus,
  Comput.\ Phys.\ Commun.\  {\bf 181} (2010) 1293--1300
  [arXiv:0912.2546 [physics. comp-ph]].
%
\bibitem{vonManteuffel:2012np}
  A.~von Manteuffel and C.~Studerus,
  {\it Reduze 2 - Distributed Feynman Integral Reduction},
  arXiv:1201. 4330 [hep-ph].
%
\bibitem{Larin:1993tq}
S.A.~Larin,
Phys. Lett. B \textbf{303} (1993) 113--118
[hep-ph/9302240].
%
\bibitem{HAMBERG}
R.~Hamberg, {\it Second order gluonic contributions to physical quantities}, PhD Thesis, U. Leiden, 
1991.
%
\bibitem{Bierenbaum:2007qe}
I.~Bierenbaum, J.~Bl\"umlein and S.~Klein,
Nucl. Phys. B \textbf{780} (2007)  40--75
[hep-ph/0703285].
%
\bibitem{Ablinger:2012qm}
J.~Ablinger, J.~Bl\"umlein, A.~Hasselhuhn, S.~Klein, C.~Schneider and F.~Wissbrock,
Nucl. Phys. B \textbf{864} (2012) 52--84
[arXiv:1206.2252 [hep-ph]].
%
\bibitem{Ablinger:2015tua}
J.~Ablinger, A.~Behring, J.~Bl{\"u}mlein, A.~De Freitas, A.~von Manteuffel and C.~Schneider,
Comput. Phys. Commun. \textbf{202} (2016) 33--112
[arXiv:1509.08324 [hep-ph]].
%
\bibitem{Bierenbaum:2008yu}
I.~Bierenbaum, J.~Blumlein, S.~Klein and C.~Schneider,
Nucl. Phys. B \textbf{803} (2008) 1--41
[arXiv:0803.0273 [hep-ph]].
%
\bibitem{Ablinger:2010ty}
  J.~Ablinger, J.~Bl\"umlein, S.~Klein, C.~Schneider and F.~Wi\ss{}brock,
  Nucl.\ Phys.\ B {\bf 844} (2011) 26--54
  [arXiv:1008.3347 [hep-ph]].
%
\bibitem{Karr:1981}
M.~Karr, 
{J.~ACM} {\bf 28} (1981) 305--350.
%
\bibitem{Bron:00}
M.~Bronstein,
{ J.~Symbolic Comput.} {\bf 29} (2000) no.~6, 841--877.
%
\bibitem{Schneider:01}
C.~Schneider, {\it Symbolic Summation in Difference Fields\/}, Ph.D. Thesis
RISC, Johannes Kepler University, Linz technical report 01--17 (2001).
%
\bibitem{Schneider:04a}
C.~Schneider, 
An. Univ. Timisoara Ser. Mat.-Inform. {\bf 42} (2004) 163--179.
%
\bibitem{Schneider:05a}
C.~Schneider, 
{J.cDiffer. Equations Appl.} {\bf 11} (2005) 799--821.
%
\bibitem{Schneider:05b}
C.~Schneider, 
Appl. Algebra Engrg. Comm. Comput. {\bf 16} (1), (2005) 1--32.
%
\bibitem{Schneider:07d}
C.~Schneider, 
J. Algebra Appl. {\bf 6} (2007) 415--441.
%
\bibitem{Schneider:2009rcr}
  C.~Schneider,
  Clay Math.\ Proc.\  {\bf 12} (2010) 285--308
  [arXiv:0904.2323 [cs.SC]].
%
\bibitem{Schneider:10c}
C.~Schneider,
{Ann. Comb.} {\bf 14} (2010) 533--552 [arXiv:0808.2596 [cs.SC]].
%
\bibitem{Schneider:15a}
 C.~Schneider, in: 
{\sf Computer Algebra and Polynomials, Applications of
  Algebra and Number Theory}, J.~Gutierrez, J.~Schicho, M.~Weimann (eds.),
  Lecture Notes in Computer Science (LNCS) 8942 (2015) 157--191
  [arXiv:1307.7887 [cs.SC]].
%
\bibitem{Schneider:08d}
C.~Schneider, 
J. Symb. Comput. {\bf 72} (2016) 
82--127 [arXiv:1408.2776 [cs.SC]].
%
\bibitem{Schneider:2017}
S.A. Abramov, M. Bronstein, M. Petkov{\v s}ek, Carsten Schneider,
J. Symb. Comput. {\bf 107} (2021)
23--66
[arXiv:2005.04944 [cs.SC]].
%
\bibitem{Blumlein:2017dxp}
J.~Bl{\"u}mlein and C.~Schneider,
Phys. Lett. B \textbf{771} (2017) 31--36
[arXiv:1701.04614 [hep-ph]].
%
\bibitem{GUESS}
M.~Kauers, {\it Guessing Handbook}, JKU Linz, Technical Report RISC 09--07.
%
\bibitem{Blumlein:2009tj}
  J.~Bl\"umlein, M.~Kauers, S.~Klein and C.~Schneider,
  Comput.\ Phys.\ Commun.\  {\bf 180} (2009) 2143--2165
  [arXiv: 0902.4091 [hep-ph]].
%
\bibitem{SAGE}
Sage, {\tt http://www.sagemath.org/}
%
\bibitem{GSAGE}
M.~Kauers, M.~Jaroschek, and F.~Johansson, 
in:
{\sf Computer Algebra and Polynomials},
Editors: J.~Gutierrez, J.~Schicho, Josef, M.~Weimann, Eds..
Lecture Notes in Computer Science {\bf 8942} (Springer, Berlin, 2015) 105--125
[arXiv:1306.4263 [cs.SC]].
%
\bibitem{SIG1}
 C.~Schneider,
{S\'em.~Lothar. Combin.\/} {\bf 56} (2007) 1--36  
 article B56b.
%
\bibitem{SIG2}
C.~Schneider, 
in:~{{\sf Computer
Algebra in Quantum Field Theory: Integration,
  Summation and Special Functions}\/} Texts and Monographs in Symbolic
  Computation eds. C.~Schneider and J.~Bl\"umlein  (Springer, Wien, 2013) 325--360
  [arXiv:1304.4134 [cs.SC]].
%
\bibitem{Blumlein:1996vs}
J.~Bl\"umlein and N.~Kochelev,
Nucl. Phys. B \textbf{498} (1997) 285--309
[hep-ph/9612318].
%
\bibitem{Ablinger:2022wbb}
J.~Ablinger, A.~Behring, J.~Bl\"umlein, A.~De Freitas, A.~Goedicke, A.~von Manteuffel, 
C.~Schneider and K.~Sch\"onwald,
JHEP \textbf{12} (2022) 134
[arXiv:2211.05462 [hep-ph]].
%
\bibitem{Blumlein:2009cf}
J.~Bl\"umlein, D.J.~Broadhurst and J.A.M.~Vermaseren,
Comput. Phys. Commun. \textbf{181} (2010) 582--625
[arXiv: 0907.2557 [math-ph]].
%
\bibitem{Ablinger:2013eba}
J.~Ablinger, J.~Bl\"umlein and C.~Schneider,
J. Phys. Conf. Ser. \textbf{523} (2014) 012060
[arXiv:1310.5645 [math-ph]].
%
\bibitem{Ablinger:2014rba}
J.~Ablinger,
PoS (LL2014)  019
[arXiv:1407.6180 [cs.SC]].
%
\bibitem{Ablinger:2010kw}
 J.~Ablinger,
  {\it A Computer Algebra Toolbox for Harmonic Sums Related to Particle Physics},
  Diploma Thesis, JKU Linz, 2009,
  arXiv:1011.1176[math-ph].
%
\bibitem{Ablinger:2013hcp}
J.~Ablinger,
{\it Computer Algebra Algorithms for Special Functions in
  Particle Physics}, Ph.D. Thesis, Linz U. (2012) arXiv:1305.0687[math-ph].
%
\bibitem{ALL2016}
J. Ablinger, 
PoS (LL2016) 067.
%
\bibitem{Ablinger:2015gdg}
J.~Ablinger,
Exper. Math. \textbf{26} (2016) no.1, 62--71
[arXiv:1507.01703 [math.NT]].
%
\bibitem{Ablinger:2018cja}
J.~Ablinger,
PoS (RADCOR2017) 001
[arXiv:1801.01039 [cs.SC]];
%
\bibitem{Ablinger:2019mkx}
J.~Ablinger,
{\it Discovering and Proving Infinite Pochhammer Sum Identities},
arXiv:1902.11001 [math.CO];
%
\bibitem{ALL2018}
J. Ablinger,
PoS (LL2018) 063.
%
\bibitem{Blumlein:2009ta}
  J.~Bl\"umlein,
  Comput.\ Phys.\ Commun.\ {\bf 180} (2009) 2218--2249
  [arXiv:0901.3106 [hep-ph]]. 
%
\bibitem{Ablinger:2013jta}
J.~Ablinger and J.~Bl\"umlein, 
in: {\sf Computer Algebra in Quantum Field Theory: Integration, Summation and
Special Fuctions}, (Springer, Wien, 2013), eds. C.~Schneider and J. Bl\"umlein, 1--32,
[arXiv:1304.7071 [math-ph]].
%
\bibitem{Ablinger:2014yaa}
J.~Ablinger, J.~Bl{\"u}mlein, C.~Raab, C.~Schneider and F.~Wi{\ss}brock,
Nucl. Phys. B \textbf{885} (2014), 409-447
[arXiv:1403.1137 [hep-ph]].
%
\bibitem{Ablinger:2018zwz}
J.~Ablinger, J.~Bl{\"u}mlein, P.~Marquard, N.~Rana and C.~Schneider,
Nucl. Phys. B \textbf{939} (2019) 253--291
[arXiv: 1810.12261 [hep-ph]].
%
\bibitem{Maier:2017ypu}
A.~Maier and P.~Marquard,
Phys. Rev. D \textbf{97} (2018) no.5, 056016
[arXiv:1710.03724 [hep-ph]].
%
\bibitem{Fael:2021kyg}
M.~Fael, F.~Lange, K.~Sch{\"o}nwald and M.~Steinhauser,
JHEP \textbf{09} (2021) 152
[arXiv:2106.05296 [hep-ph]].
%
\bibitem{Fael:2022miw}
M.~Fael, F.~Lange, K.~Sch\"onwald and M.~Steinhauser,
Phys. Rev. D \textbf{106} (2022) no.3, 034029
[arXiv:2207.00027 [hep-ph]].
%
\bibitem{Ablinger:2018brx}
J.~Ablinger, J.~Bl\"umlein, A.~De Freitas, A.~Goedicke, C.~Schneider and K.~Sch\"onwald,
Nucl. Phys. B \textbf{932} (2018) 129--240
[arXiv:1804.02226 [hep-ph]].
%
\bibitem{Ablinger:2020snj}
J.~Ablinger, J.~Bl{\"u}mlein, A.~De Freitas, A.~Goedicke, M.~Saragnese, C.~Schneider and 
K.~Sch{\"o}nwald,
Nucl. Phys. B \textbf{955} (2020) 115059
[arXiv:2004.08916 [hep-ph]].
%
\bibitem{Ablinger:2017xml}
J.~Ablinger, J.~Bl{\"u}mlein, A.~De Freitas, C.~Schneider and K.~Sch{\"o}nwald,
Nucl. Phys. B \textbf{927} (2018) 339--367
[arXiv:1711.06717 [hep-ph]]
%
\bibitem{Ablinger:2019gpu}
J.~Ablinger, J.~Bl{\"u}mlein, A.~De Freitas, M.~Saragnese, C.~Schneider and K.~Sch{\"o}nwald,
Nucl. Phys. B \textbf{952} (2020) 114916
[arXiv:1911.11630 [hep-ph]].
%
\bibitem{AITKEN}
A. Aitken, 
Proc. Royal Society of Edinburgh {\bf 46} (1926) 289--305.
%
\bibitem{Blumlein:2018cms}
J.~Bl{\"u}mlein and C.~Schneider,
Int. J. Mod. Phys. A \textbf{33} (2018) no.17, 1830015
[arXiv:1809.02889 [hep-ph]].
%
\bibitem{Blumlein:2021ynm}
J.~Bl{\"u}mlein and C.~Schneider, eds.,
{\sf Anti-Differentiation and the Calculation of Feynman Amplitudes},
doi:10.1007/978-3-030-80219-6
(Springer, Berlin, 2021).
%
\bibitem{Witten:1975bh}
E.~Witten,
Nucl. Phys. B \textbf{104} (1976) 445--476.
%
\bibitem{Babcock:1977fi}
J.~Babcock, D.W.~Sivers and S.~Wolfram,
Phys. Rev. D \textbf{18} (1978) 162--181.
%
\bibitem{Shifman:1977yb}
M.A.~Shifman, A.I.~Vainshtein and V.I.~Zakharov,
Nucl. Phys. B \textbf{136} (1978) 157--176.
%
\bibitem{Leveille:1978px}
J.P.~Leveille and T.J.~Weiler,
Nucl. Phys. B \textbf{147} (1979) 147--173.
%
\bibitem{Watson:1981ce}
  A.D.~Watson,
  {Z.\ Phys.}\ C {\bf 12} (1982) 123--125.
%
\bibitem{Hekhorn:2018ywm}
F.~Hekhorn and M.~Stratmann,
Phys. Rev. D \textbf{98} (2018) no.1, 014018
[arXiv:1805.09026 [hep-ph]].
%
\bibitem{Bierenbaum:2022biv}
I.~Bierenbaum, J.~Bl\"umlein, A.~De Freitas, A.~Goedicke, S.~Klein and K.~Sch\"onwald,
Nucl. Phys. B \textbf{988} (2023) 116114
[arXiv:2211.15337 [hep-ph]].
%
\bibitem{Bierenbaum:2009zt}
  I.~Bierenbaum, J.~Bl\"umlein and S.~Klein,
  Phys.\ Lett.\ B {\bf 672} (2009) 401--406
  [arXiv:0901.0669 [hep-ph]].
%
\bibitem{Kawamura:2012cr}
H.~Kawamura, N.~A.~Lo Presti, S.~Moch and A.~Vogt,
Nucl. Phys. B \textbf{864} (2012) 399--468
[arXiv: 1205.5727 [hep-ph]].
%
\bibitem{Behring:2014eya}
  A.~Behring, I.~Bierenbaum, J.~Bl\"umlein, A.~De Freitas, S.~Klein and F.~Wi\ss{}brock,
  Eur.\ Phys.\ J.\ C {\bf 74} (2014) 9,  3033
  [arXiv:1403.6356 [hep-ph]].
%
\bibitem{Blumlein:2021xlc}
J.~Bl{\"u}mlein, A.~De Freitas, M.~Saragnese, C.~Schneider and K.~Sch{\"o}nwald,
Phys. Rev. D \textbf{104} (2021) no.3, 034030
[arXiv:2105.09572 [hep-ph]].
%
\bibitem{Blumlein:2006mh}
J.~Bl\"umlein, A.~De Freitas, W.L.~van Neerven and S.~Klein,
Nucl. Phys. B \textbf{755} (2006) 272--285
[hep-ph/0608024].
%
\bibitem{Ablinger:2014vwa}
  J.~Ablinger, A.~Behring, J.~Bl\"umlein, A.~De Freitas, A.~Hasselhuhn, A.~von Manteuffel, 
  M.~Round, C.~Schneider, and F.~Wi\ss{}brock,
  Nucl.\ Phys.\ B {\bf 886} (2014) 733--823
  [arXiv:1406.4654 [hep-ph]].
%
\bibitem{Ablinger:2014nga}
  J.~Ablinger, A.~Behring, J.~Bl\"umlein, A.~De Freitas, A.~von Manteuffel and C.~Schneider,
  Nucl.\ Phys.\ B {\bf 890} (2014) 48--151
  [arXiv:1409.1135 [hep-ph]].
%
\bibitem{Ablinger:2024xtt} 
J.~Ablinger, A.~Behring, J.~Bl\"umlein, A.~De Freitas, A.~von Manteuffel, 
C.~Schneider and K.~Sch\"onwald,
Phys. Lett. B \textbf{854} (2024) 138713
[arXiv:2403.00513 [hep-ph]].
%
\bibitem{Ablinger:2019etw}
J.~Ablinger, A.~Behring, J.~Bl\"umlein, A.~De Freitas, A.~von Manteuffel, C.~Schneider and K.~Sch\"onwald,
Nucl. Phys. B \textbf{953} (2020) 114945
[arXiv:1912.02536 [hep-ph]].
%
\bibitem{Behring:2015zaa}
A.~Behring, J.~Bl{\"u}mlein, A.~De Freitas, A.~von Manteuffel and C.~Schneider,
Nucl. Phys. B \textbf{897} (2015) 612--644
[arXiv:1504.08217 [hep-ph]]
%
\bibitem{Blumlein:2024euz}
J.~Bl\"umlein and M.~Saragnese,
Phys. Rev. D \textbf{110} (2024) no.3, 034006
[arXiv:2405.17252 [hep-ph]].
%
\bibitem{Gluck:1996ve}
M.~Gl\"uck, S.~Kretzer and E.~Reya,
Phys. Lett. B \textbf{380} (1996) 171--176
[erratum: Phys. Lett. B \textbf{405} (1997), 391]
[arXiv:hep-ph/9603304 [hep-ph]].
%
\bibitem{Blumlein:2011zu}
J.~Bl\"umlein, A.~Hasselhuhn, P.~Kovacikova and S.~Moch,
Phys. Lett. B \textbf{700} (2011)  294--304
[arXiv:1104.3449 [hep-ph]].
%
\bibitem{Buza:1997mg}
M.~Buza and W.~L.~van Neerven,
Nucl. Phys. B \textbf{500} (1997) 301--324
[hep-ph/9702242].
%
\bibitem{Blumlein:2014fqa}
J.~Bl\"umlein, A.~Hasselhuhn and T.~Pfoh,
Nucl. Phys. B \textbf{881} (2014) 1-41
[arXiv:1401.4352 [hep-ph]].
%
\bibitem{Behring:2015roa}
A.~Behring, J.~Bl{\"u}mlein, A.~De Freitas, A.~Hasselhuhn, A.~von Manteuffel and C.~Schneider,
Phys. Rev. D \textbf{92} (2015) no.11, 114005
[arXiv:1508.01449 [hep-ph]].
%
\bibitem{Behring:2016hpa}
A.~Behring, J.~Bl{\"u}mlein, G.~Falcioni, A.~De Freitas, A.~von Manteuffel and C.~Schneider,
Phys. Rev. D \textbf{94} (2016) no.11, 114006
[arXiv:1609.06255 [hep-ph]].
%
\bibitem{Karlberg:2025hxk}
A.~Karlberg, P.~Nason, G.~Salam, G.~Zanderighi and F.~Dreyer,
{\it HOPPET v2 release note},
[arXiv:2510.09310 [hep-ph]]
%
\bibitem{Ablinger:2014lka}
  J.~Ablinger, J.~Bl\"umlein, A.~De Freitas, A.~Hasselhuhn, A.~von Manteuffel, M.~Round, C.~Schneider and 
  F.~Wi\ss{}brock,
  Nucl.\ Phys.\ B {\bf 882} (2014) 263--288
  [arXiv:1402.0359 [hep-ph]].
%
\bibitem{Behring:2021asx}
A.~Behring, J.~Bl\"umlein, A.~De Freitas, A.~von Manteuffel, K.~Sch\"onwald and C.~Schneider,
Nucl. Phys. B \textbf{964} (2021) 115331
[arXiv:2101.05733 [hep-ph]].
%
\bibitem{Blumlein:2009rg}
J.~Bl\"umlein, S.~Klein and B.~T\"odtli,
Phys. Rev. D \textbf{80} (2009) 094010
[arXiv:0909.1547 [hep-ph]].
%
\bibitem{Catani:1990eg}
S.~Catani, M.~Ciafaloni and F.~Hautmann,
Nucl. Phys. B \textbf{366} (1991) 135--188.
%
\bibitem{Alekhin:2017kpj}
S.~Alekhin, J.~Bl{\"u}mlein, S.~Moch and R.~Placakyte,
Phys. Rev. D \textbf{96} (2017) no.1, 014011
[arXiv:1701.05838 [hep-ph]].
%
\bibitem{Blumlein:1997em}
J.~Bl\"umlein and A.~Vogt,
Phys. Rev. D \textbf{58} (1998) 014020 [arXiv:hep-ph/9712546 [hep-ph]].
%
\bibitem{Blumlein:1998mg}
J.~Bl\"umlein and W.~L.~van Neerven,
Phys. Lett. B \textbf{450} (1999) 412--416
[hep-ph/9811519].
%
\bibitem{Blumlein:1999ev}
J.~Bl\"umlein,
Lect. Notes Phys. \textbf{546} (2000) 42--57
[arXiv:hep-ph/9909449 [hep-ph]].
%
\bibitem{Blumlein:2018jfm}
J.~Bl{\"u}mlein, A.~De Freitas, C.~Schneider and K.~Sch{\"o}nwald,
Phys. Lett. B \textbf{782} (2018) 362--366
[arXiv:1804.03129 [hep-ph]].
%
\bibitem{Boer:2011fh}
D.~Boer, et al.
{\it Gluons and the quark sea at high energies: Distributions, polarization, tomography},
[arXiv: 1108.1713 [nucl-th]].
%
\bibitem{Blumlein:2021lmf}
J.~Bl\"umlein and M.~Saragnese,
Phys. Lett. B \textbf{820} (2021) 136589
[arXiv:2107.01293 [hep-ph]].
%
\bibitem{Blumlein:1987xk}
J.~Bl\"umlein, M.~Klein, T.~Naumann and T.~Riemann,
{\it Structure Functions, Quark Distributions and $\Lambda_{QCD}$ at HERA},
PHE-88-01,
       Proc. of the HERA Workshop, Hamburg, October 1987,
                                                     ed.~R.D.~Peccei,
       Vol. {\bf 1}, p.~67--105.
%
\bibitem{DESYPRC}
T.~Alexopoulos, et al., 
{\it Electron-Deuteron Scattering with HERA: a Letter of Intent, submitted to the DESY
PRC}, DESY-03-194, PRC 03/02, H1-04/03-609.
%
\bibitem{STAUDE}
C.~Guyot, A.~Milsztajn, A.~Oraou, A.~Staudte, K.M.~Teichert, and M.~Virchaux, {\it
A new fixed[target experiment for precise tests of QCD}, CERN/SPSC 89/56, March 14 1989.
%
\bibitem{Matiounine:1998re}
Y.~Matiounine, J.~Smith and W.L.~van Neerven,
Phys. Rev. D \textbf{58} (1998) 076002
[hep-ph/9803439].
%
\bibitem{Georgi:1976ve}
H.~Georgi and H.~D.~Politzer,
Phys. Rev. D \textbf{14} (1976) 1829--1848.
%
\bibitem{Schienbein:2007gr}
I.~Schienbein, V.~A.~Radescu, G.~P.~Zeller, M.~E.~Christy, C.~E.~Keppel, K.~S.~McFarland, 
W.~Melnitchouk, F.~I.~Olness, M.~H.~Reno and F.~Steffens, \textit{et al.}
J. Phys. G \textbf{35} (2008) 053101
[arXiv:0709.1775 [hep-ph]].
%
\bibitem{Piccione:1997zh}
A.~Piccione and G.~Ridolfi,
Nucl. Phys. B \textbf{513} (1998) 301--316
[hep-ph/9707478].
%
\bibitem{Blumlein:1998nv}
J.~Bl\"umlein and A.~Tkabladze,
Nucl. Phys. B \textbf{553} (1999) 427--464
[hep-ph/9812478].
%
\bibitem{Christy:2012tk}
M.E.~Christy, J.~Bl\"umlein and H.~B\"ottcher,
{\it Unfolding of target mass contributions from inclusive proton structure function 
data},
[arXiv:1201.0576 [hep-ph]].
%
\bibitem{Kwiatkowski:1990es}
A.~Kwiatkowski, H.~Spiesberger and H.J.~M\"ohring,
Comput. Phys. Commun. \textbf{69} (1992) 155--172
%
\bibitem{Arbuzov:1995id}
A.~Arbuzov, D.Y.~Bardin, J.~Bl\"umlein, L.~Kalinovskaya and T.~Riemann,
Comput. Phys. Commun. \textbf{94} (1996) 128--184
[hep-ph/9511434].
%
\bibitem{Moch:2004pa}
  S.~Moch, J.A.M.~Vermaseren and A.~Vogt,
  Nucl.\ Phys.\ B {\bf 688} (2004) 101--134
  [hep-ph/0403192];
\textbf{691} (2004) 129--181
[hep-ph/0404111].
%
\bibitem{Ablinger:2017tan}
J.~Ablinger, A.~Behring, J.~Bl{\"u}mlein, A.~De Freitas, A.~von Manteuffel and 
C.~Schneider,
Nucl. Phys. B \textbf{922} (2017) 1-40
[arXiv:1705.01508 [hep-ph]].
%
\bibitem{Blumlein:2021enk}
J.~Bl\"umlein, P.~Marquard, C.~Schneider and K.~Sch\"onwald,
Nucl. Phys. B \textbf{971} (2021) 115542
[arXiv:2107.06267 [hep-ph]].
%
\bibitem{HADR}
C.~Anastasiou, C.~Duhr, F.~Dulat, F.~Herzog and B.~Mistlberger,
Phys. Rev. Lett. \textbf{114} (2015) 212001
[arXiv:1503.06056 [hep-ph]]. \\
B.~Mistlberger,
JHEP \textbf{05} (2018) 028
[arXiv:1802.00833 [hep-ph]]. \\
M.X.~Luo, T.Z.~Yang, H.X.~Zhu and Y.J.~Zhu,
Phys. Rev. Lett. \textbf{124} (2020)  092001
[arXiv:1912.05778 [hep-ph]];
JHEP \textbf{06} (2021) 115
[arXiv:2012.03256 [hep-ph]].\\
M.A.~Ebert, B.~Mistlberger and G.~Vita,
JHEP \textbf{09} (2020) 146
[arXiv:2006.05329 [hep-ph]].
\textbf{09} (2020) 143
[arXiv:2006.03056 [hep-ph]].\\
D.~Baranowski, A.~Behring, K.~Melnikov, L.~Tancredi and C.~Wever,
JHEP \textbf{02} (2023) 073
[arXiv:2211.05722 [hep-ph]].
%
\bibitem{Duhr:2020seh}
C.~Duhr, F.~Dulat and B.~Mistlberger,
Phys. Rev. Lett. \textbf{125} (2020) 172001
[arXiv:2001.07717 [hep-ph]];
JHEP \textbf{11} (2020) 143
[arXiv:2007.13313 [hep-ph]].
%
\bibitem{Gehrmann:2023ksf}
T.~Gehrmann, A.~von Manteuffel and T.Z.~Yang,
JHEP \textbf{04} (2023) 041
[arXiv:2302.00022 [hep-ph]].
%
\bibitem{Moch:2014sna}
S.~Moch, J.A.M.~Vermaseren and A.~Vogt,
Nucl. Phys. B \textbf{889} (2014) 351--400
[arXiv:1409.5131 [hep-ph]].
%
\bibitem{Blumlein:2021ryt}
J.~Bl\"umlein, P.~Marquard, C.~Schneider and K.~Sch\"onwald,
JHEP \textbf{01} (2022) 193
[arXiv:2111.12401 [hep-ph]].
%
\bibitem{Behring:2025avs}
A.~Behring, J.~Bl{\"u}mlein, A.~De Freitas, A.~von Manteuffel, C.~Schneider and 
K.~Sch{\"o}nwald,
{\it The heavy quark-antiquark asymmetry in the variable flavor number scheme},
[arXiv:2512.13508 [hep-ph]].
%
\bibitem{Zhu:2025gts}
Y.J.~Zhu,
{\it The N$^3$LO Twist-2 Matching of Helicity TMDs and SIDIS $q_\ast$ Spectrum},
[arXiv:2509.01655 [hep-ph]].
%
\bibitem{SPLINE}
A.~Klein and A.~Godunov, {\sf Introductory Computational Physics}, (Cambridge University 
Press, Cambridge, 2006);\\
{\tt 
https://github.com/jannisteunissen/spline\_interpolation\_fortran?tab=readme
}\\
{\tt  -ov-file}
%
\bibitem{Ruijl:2017dtg}
B.~Ruijl, T.~Ueda and J.~Vermaseren,
{\it FORM version 4.2}, [arXiv:1707.06453 [hep-ph].
%
\bibitem{AIND}
R.~Piessens, 
Angew. Informatik {\bf 9} (1973) 399--401.
%
\bibitem{LHeCStudyGroup:2012zhm}
J.~L.~Abelleira Fernandez \textit{et al.} [LHeC Study Group],
J. Phys. G \textbf{39} (2012), 075001
[arXiv:1206.2913 [physics.acc-ph]].
%
\bibitem{LHeC:2020van}
P.~Agostini \textit{et al.} [LHeC and FCC-he Study Group],
J. Phys. G \textbf{48} (2021) no.11, 110501
[arXiv:2007.14491 [hep-ex]].
\end{thebibliography}
\end{document}